\documentclass[11pt,twoside]{article}

\usepackage{asp2006}
\usepackage{fancyhdr,graphicx,caption,rotating,multirow,lscape}
\DeclareGraphicsExtensions{.eps,.eps.gz,.ps,.jpg,.tiff}

\begin{document}

\title{Early Results from the KELT Transit Survey}  

\author{Joshua Pepper, Richard Pogge, D.~L. DePoy, J.~L. Marshall, Kris Stanek}
\affil{Department of Astronomy, The Ohio State University, 140 West 18th Avenue, Columbus, OH, 43210-1173}
\author{Amelia Stutz}  
\affil{Department of Astronomy, Univeristy of Arizona, 933 N Cherry Ave, Tucson AZ 85721-0065}
\author{Mark Trueblood, Pat Trueblood}  
\affil{Winer Observatory, P.\ O.\ Box 797, Sonoita, AZ 85637-0797}

\begin{abstract}
The Kilodegree Extremely Little Telescope (KELT) project is a small-aperture transit
survey of bright stars. The project has completed commissioning runs searching for transits in the
Hyades and Praesepe, and is well into a multi-year survey of a large portion of the Northern 
Hemisphere.  Here we describe the setup of the telescope and discuss the early data.
\end{abstract}

There is great scientific potential in the discovery of transiting planets with 
bright host stars.  Follow-up observation of such systems can determine  
properties of the planets that are not measurable in non-transiting systems, or 
in transiting systems where the host stars are too faint.  The Kilodegree Extremely Little 
Telescope (KELT) project is a wide field, small-aperture survey for such 
transits, based on a theoretical optimization of all-sky transit 
searches \citep{pep03}.  It is similar to the other small-telescope transit surveys with wide 
fields \citep{alonso04,mc06,bakos04,acc06}, but it surveys a larger area of sky 
than the other surveys.  KELT is designed to target stars with 
magnitudes $8 < V < 10$; a range fainter than the stars observed in radial-velocity 
surveys but brighter than those targeted by most existing transit surveys.  KELT
consists of a single automated telescope located at the remote telescope hosting 
site Winer Observatory in Sonoita, AZ.  The primary KELT mission is a survey of a 
strip of sky 26 degrees wide at $\delta = +32^\circ$, broken 
into 13 equally-spaced fields.  The telescope began operations in October 2004, and 
spent two commissioning runs observing the open clusters Praesepe and the 
Hyades.  We then began observing the 13 survey fields, which together cover 
about 25\% of the Northern sky.

We have been gathering data for the past two years, with gaps due to the need 
to shut down during the summer monsoons, and also due to periodic equipment 
problems.  The telescope is currently operating well, and returning data 
regularly.  In this proceeding, we briefly describe the telescope instrumentation, 
the survey area and observing strategy, and we discuss the telescope performance 
and show sample lightcurves. 

\section{Instrumentation} 

\begin{figure}[!ht]
\centering
\includegraphics[angle=0,width=12cm]{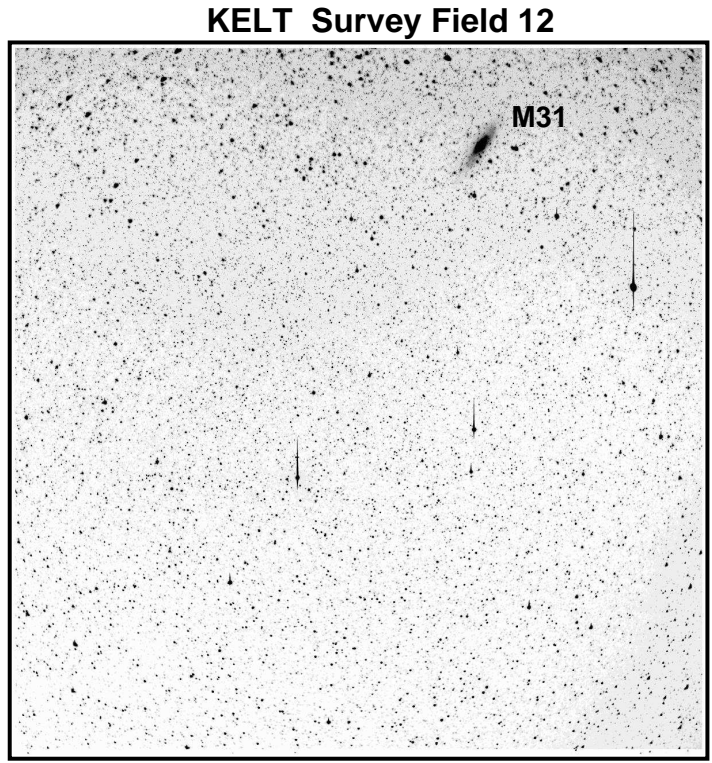}
\vspace{-1in}
\caption{Image of one $26^{\circ} \times 26^{\circ}$ survey field from the
KELT camera.  The extended object at the upper right is M31. \label{fig:m31}}
\vspace{1in}
\end{figure}

KELT employs an Apogee AP16E thermoelectrically cooled CCD camera.  This 
camera uses the Kodak KAF-16801E front-side illuminated CCD 
with $4096 \times 4096$ $9\mu m$ pixels (36.88 $\times$ 36.88\,mm detector 
area).  We use two different lenses with KELT.  For the wide-angle survey mode,
we use a Mamiya 645 80\,mm f/1.9 medium-format manual-focus lens with a
42\,mm aperture.  This lens provides a roughly 23\arcsec\,pix$^{-1}$ image scale 
and a $26^\circ \times 26^\circ$ field of view, with vignetting at the corners, so 
the effective clear field of view is circular.  Figure \ref{fig:m31} shows a sample 
image of a survey field.  We selected the field with galaxy M31 to give perspective 
on the size of our fields.  To provide a narrow-angle campaign mode used for the 
Hyades and Praesepe observations, we use a Mamiya 645 200\,mm f/2.8 APO manual-focus 
telephoto lens with a 71\,mm aperture.  This provides a roughly 9.5\arcsec\,pix$^{-1}$ image 
scale and effective 10\fdg8$\times$10\fdg8 field of view.

To reject the mostly-blue background sky without greatly diminishing the
sensitivity to stars (which are mostly redder than the night sky), we
use a Kodak Wratten \#8 red-pass filter with a 50\% transmission point
at $\sim$490\,nm (the filter looks yellow to the eye).  The filter is
mounted in front of the KELT lens during operations.  

The optical assembly (camera+lens+filter) is mounted on a Paramount ME
Robotic Telescope Mount manufactured by Software
Bisque.  The Paramount is a research-grade German Equatorial Mount designed 
specifically for robotic operation with integrated telescope and camera 
control.  The CCD camera and mount are controlled by a PC computer located 
at the observing site that runs Windows XP Pro and the Bisque Observatory 
Software Suite from Software Bisque.  

\section{Survey Area and Strategy}

The main targets for KELT are a series of fields, all at $\delta = +32^\circ$, the 
latitude of Winer Observatory.  The fields are 
spaced equally in Right Ascension, and we tile between two fields at any given 
time, observing each once before slewing to the next field, yielding a cadence 
on a single field of about six minutes, with an exposure time of 2.5 minutes 
and 40 seconds for readout and slewing.  Over the course of a year, we take 
about the same number of images from each field.  Since beginning operations, we 
have taken over 40,000 images of our 13 survey fields.  We are still building 
the software pipeline for the data and gathering images from certain fields.  By 
spring 2007, we expect to begin searching our data for transits.

Each survey field is about $500^2$ degrees, and together they fill about 25\% of 
the Northern sky.  This amount of coverage and the magnitude range of our target 
stars set KELT apart from similar wide-angle surveys.  This setup allows KELT to
 search for the most scientifically interesting transits over the largest part of 
the sky.  Based on estimates from \citet{pep03}, we expect to find $\sim 4$ 
transiting planets in these fields.

\section{Performance}

The critical threshold for sensitivity to transits is generally taken to be an RMS 
of less than 1\%, which corresponds to the transit by a Jupiter-sized planet of 
a solar-type star.  In order to achieve this level of precision, we employ image 
subtraction using the ISIS package \citep{alard00}.  Difference imaging is often 
difficult to implement in general, and is especially prone to complications with very 
large fields of view.  After overcoming various problems, we were able to successfully 
implement ISIS.  Among transit surveys, KELT has the largest field of view for which image subtraction has 
been successfully employed.

\begin{figure}[!ht]
\centering
\includegraphics[angle=270,width=15cm]{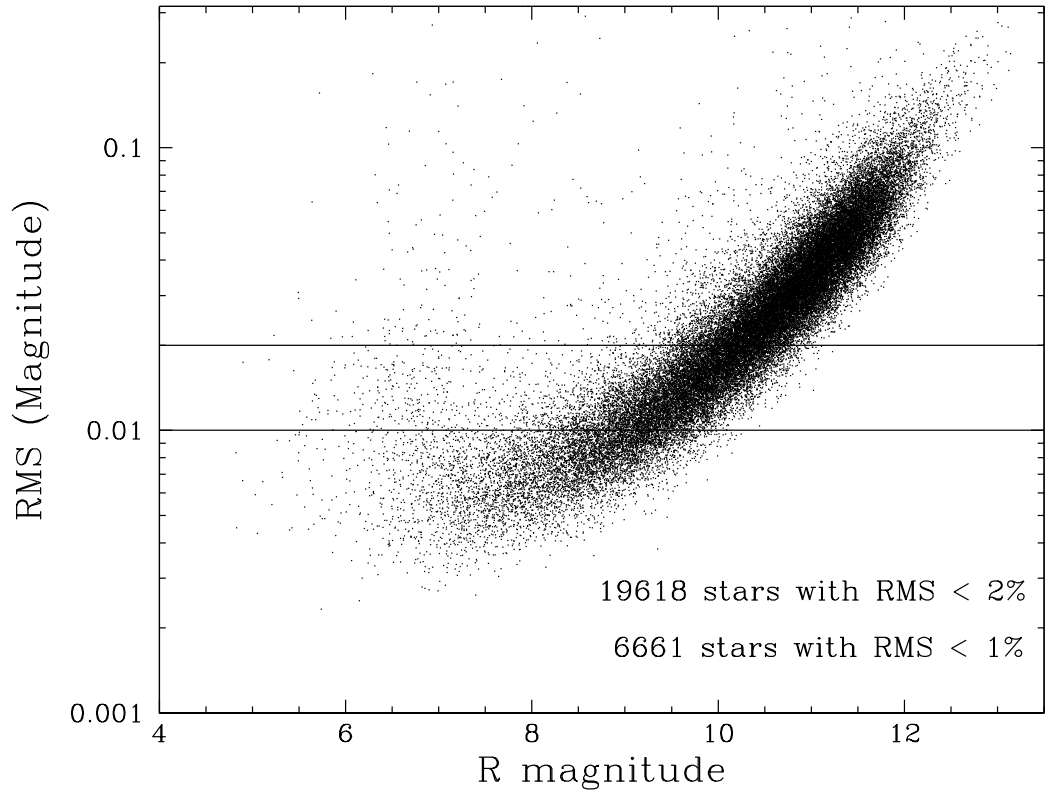}
\caption{RMS magnitude scatter of a time-series of magnitude measurements of one of the survey fields.  This plot shows the RMS for 33 images of 73,000 stars over one night of observing, with no detrending applied.\label{fig:rms}}
\end{figure}

Figure \ref{fig:rms} shows the RMS for one of our survey fields 
from one night of observing.  There are over 6000 stars in the field showing sub-1\% 
variation, which would make it possible to detect transits in their lightcurves.  A more robust 
measure of transit sensitivity would require computing RMS over the entire course of 
observations of the field, but would also require the application of a detrending 
algorithm, such as SYSREM \citep{tamuz05}, that would remove nightly atmospheric 
and airmass variations.  We are still testing detrending algorithms with our data, 
and thus do not yet have a complete analysis for our long-term RMS sensitivity.  However, 
the quality of the single-night lightcurves demonstrate the potential of the data.  A full 
description of the performance of the telescope will appear in an upcoming paper, which 
will also provide a detailed discussion of the telescope instrumentation and control and 
data handling procedures \citep{pepper06}.

\section{Example Lightcurves}

\begin{figure}[!ht]
\centering
\includegraphics[angle=0,width=9cm]{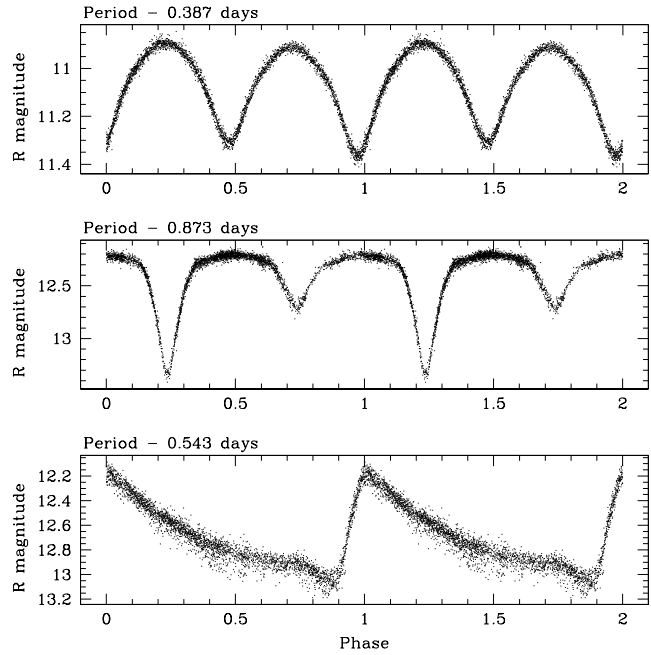}
\vspace{-0.2in}
\caption{Examples of newly discovered variable stars in Praesepe.  Each light curve contains over 3000 points. \label{fig:vars}}
\end{figure}

\begin{figure}[!ht]
\centering
\includegraphics[angle=0,width=9cm]{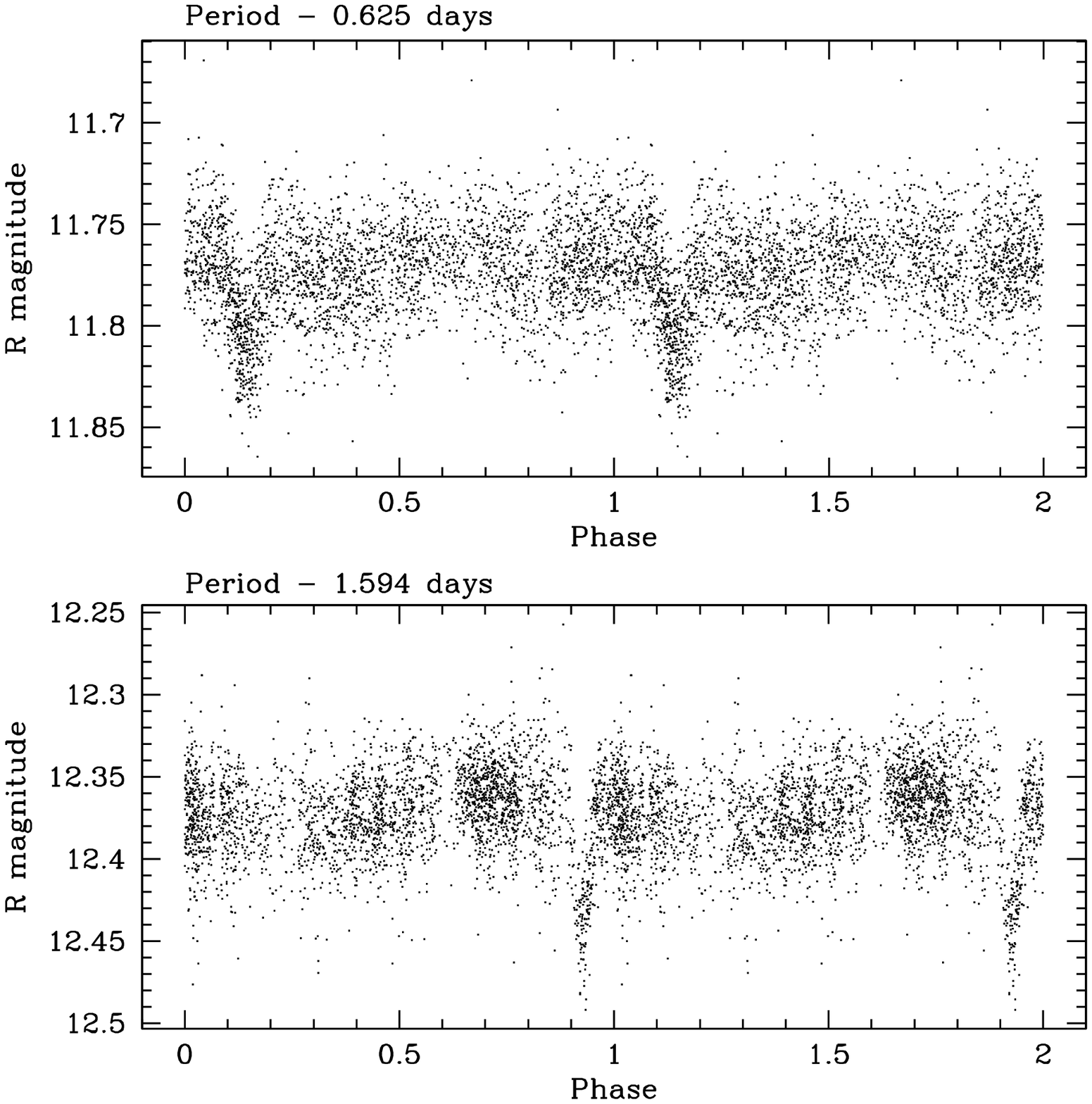}
\vspace{-0.1in}
\caption{Transit-like light curves from KELT observations of Praesepe.  These objects are probably not planets, but demonstrate the ability of KELT to detect transits.\label{fig:dips}}
\end{figure}

Along with the survey data taken to date, we have been analyzing the commissioning 
data on Praesepe, which consists of over 3,000 observations taken on 34 nights over the course of 74 
days, using the 200\,mm lens with the 10\fdg8$\times$10\fdg8 fields of view.  We have found 
a large number of variable stars, of which we show three examples 
in Figure \ref{fig:vars}.  We have also begun searching the Praesepe data for transits.  We 
are still sorting through the data, but we have found two-transit-like curves, shown 
in Figure \ref{fig:dips}.  These signals are probably not planetary transits, based on their 
depths and periods, but they demonstrate our ability to detect transits with the KELT data.  We 
will be publishing an upcoming paper describing the analysis of the Praesepe data for 
variable stars and transits \citep{pepper07}. 

\vspace{0.2in}

\vspace{0.5in}
\acknowledgements 
Thanks to Scott Gaudi for helpful comments.  Thanks also to the Local Organizing Committee of 
this Workshop, Cristina Afonso, David Weldrake, and Maria Janssen-Bennynck.  This 
work was supported by the National Aeronautics and Space Administration under Grant No. NNG04GO70G issued 
through the Origins of Solar Systems program.

\end{document}